# A novel approach to process TRISO nuclear fuel using plasma-aided chemistry

T. Chemnitz[1], C. Reiter[1,3], F. Kraus[2] and D. Novog[3]

1) Forschungs-Neutronenquelle Heinz Maier-Leibnitz, TU München, Lichtenbergstraße 1, 85748 Garching, Germany

2) Fachbereich Chemie, Philipps-Universität Marburg, Hans-Meerwein-Str. 4, 35032 Marburg, Germany

3) Department for Engineering Physics, McMaster University, 1280 Main Street West, Hamilton, ON L8S 4L8, Canada

**Abstract**

This paper provides a unique and to the best of our knowledge first-of-a-kind attempt to develop chemical processes that may contribute to the volume reduction of SMR TRISO-based fuels and aims at the eventual ability to reprocess the spent fuel. To this end, the etching behavior of two materials, silicon carbide, SiC and pyrolytic carbon, PyC, that are generally used for the different barrier layers of a TRISO particle has been investigated. Either F/NF$_x$ or O radicals were used as etching agents and were obtained from NF$_3$ and molecular O$_2$, respectively, using a microwave plasma generated in a remote plasma source RPS. Laser heating of the sample materials of up to 1200 °C allowed for determination of etching rates. The results of these experiments show that chemical processing of TRISO particles via plasma-assisted etching is possible and complete removal of the encapsulation and TRISO layers can be achieved. Additional research on the waste streams and off-gases, in particular the possibility of introducing intermediate steps to reduce the CO$_2$ formed in the chemical reactions is needed.

## 1. Introduction

The tristructural-isotropic (TRISO) nuclear fuel was originally developed in the United States and the United Kingdom in the 1960s. Thereby, the main focus was the development of a versatile fuel able to withstand the high fuel temperatures in gas-cooled reactors as well as superior performance in accident scenarios. Recently, TRISO fuel has regained significant attention in both the Accident Tolerant Fuel (ATF) applications as well as in Small Modular Reactors (SMRs) [6]. The use of TRISO as a nuclear fuel is not a completely new concept and was intensively investigated in the past, e.g., in the German pebble bed reactor HTR (High Temperature Reactor) of Arbeitsgemeinschaft Versuchsreaktor Jülich (AVR) [8]. The United States performed an extensive fuel qualification program, which laid the basis for TRISO being a viable fuel form for SMRs and as Accident Tolerant Fuel in light water reactors [5]. This period of development proved that TRISO fuel could operate reliably at temperatures and efficiency levels unattainable with conventional nuclear fuel.

TRISO fuel also plays an important role in the current push for the deployment of SMRs as it enables flexible and compact reactor designs and can tolerate both high normal operating temperatures and a wide range of accident conditions, due to the material properties and multiple barriers. The companies X-Energy and NANO Nuclear Energy Inc. are proposing TRISO fuel for their respective Micro Modular Reactor design [9, 11]. In 2021, China commissioned the HTR-PM, a modern pebble-bed high-temperature gas-cooled reactor also using TRISO fuel [13]. In South Africa, the Pebble Bed Modular

Reactor (PBMR), for which the development started in the 1990s, also regained attention and the government committed to a revival of the PBMR [12].

The center of each TRISO particle is a solid, spherical kernel consisting of a uranium compound serving as nuclear fuel. Preferred compounds are uranium dioxide, $UO_2$, uranium mononitride, UN, or a uranium oxide-carbide mixture, commonly abbreviated as UCO, although not being a chemical compound of this composition. The fuel is encapsulated by a four-layered system. Going further from the inside to the outside, it consists of a layer of porous carbon acting as a plenum to ensure the retention of fission products, an inner layer of pyrolytic carbon (IPyC), a layer of SiC acting as a barrier, and a final outer layer of pyrolytic carbon (OPyC). This material selection and arrangement provides superior fission product retention up to 1600 °C for over 300 h [4]. A model of a TRISO particle is depicted in Figure 1. Since each particle measures less than 1.0 mm in diameter, hundreds or thousands of particles are embedded in a matrix material such as graphite or SiC. While the matrix material provides structural stability of the fuel system, it also serves as ultimate barrier for fission products, in case they are not retained by the SiC layer.

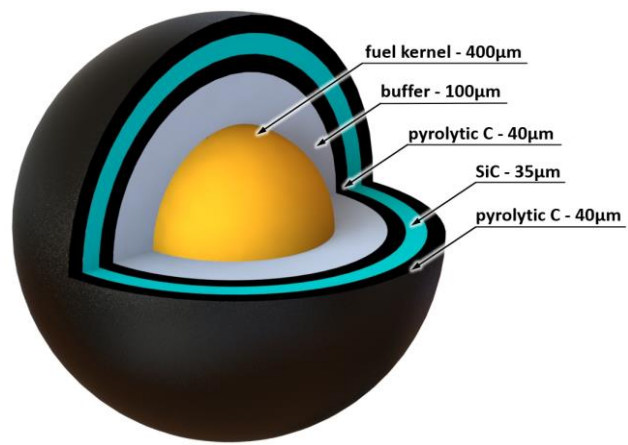

Figure 1: Schematic structure of a TRISO particle (modified from [3]).

Even though TRISO particles offer great opportunities for reactor designs, a clear backend path is not yet available. As for standard $UO_2$ fuel for LWRs, TRISO could be put at a final disposal or fed back to the fuel cycle by reprocessing of the fuel kernel. For final disposal, it is desirable to remove the graphite matrix in order to reduce the volume to be stored and at the same time to leave the SiC barrier functional to ensure a safe enclosure of the fission products [7]. For reprocessing, which would allow for a much-increased total fuel efficiency, the fuel kernel needs to be stripped of all barriers.

The reprocessing option for its part relies on mechanical processes such as cracking up the particles by grinding and treating the constituents separately. While such grinding processes are not material selective and due to the material properties might also induce significant material wear on the equipment, according to [1], they are also deemed to hardly be socially or legally acceptable today due to potential environmental emissions.

To address these challenges and to potentially open a viable reprocessing pathway for TRISO fuel, the authors propose a novel process that focuses on its dry chemical dissolution. This paper presents the first steps and results towards a dry chemical reprocessing process based on plasma-assisted etching.

## 2. Materials and methods

The compounds used for the different layers of TRISO fuel are susceptible to plasma etching using either F/NF$_x$, or O radicals. These radical species show quite a different etching behavior for SiC and PyC, thus allowing for a certain degree of specificity during the etching process. This paper thereby focuses on the coating materials of the fuel, SiC and PyC, where we prove the possibility of removal of each of these layers through separate etching experiments. The F/NF$_x$ radicals are used for the etching of the SiC layer, whereas the O radicals are used for the etching of the different carbon layers, namely the IPyC and the OPyC layers as well as the buffer layer. Depending on the matrix material, which encapsulates the TRISO particles, either fluorine or oxygen radicals may be used for its decomposition.

The proposed process uses a modified Schlenk line [7], an apparatus that allows exclusion of moisture and air. Three different process gases, Ar (4.8), NF$_3$ (4.0) and O$_2$ (3.8) can be fed into a remote plasma source (RPS) by three mass flow controllers (MFCs). The RPS is designed in such a way that the plasma exclusively burns inside the RPS. For the experiments presented here, NF$_3$ and O$_2$ serve as etching agents and dissociate in the RPS into F/NF$_x$ and O radicals, respectively. Ar on the other hand serves as a carrier gas. Directly connected to the RPS is the reaction chamber made out of Monel, into which the substrate is placed. The substrate material for its part can selectively be heated using a 100 W diode laser operating at a wavelength of 980 nm. The temperature is measured with a pyrometer, which also controls the power output of the laser. The maximum temperature depends on the substrate material and its optical properties such as reflectivity and emissivity. The reaction chamber can be cooled at the same time, thereby minimizing corrosion effects. Due to the highly corrosive nature of fluorine radicals, all metal components are made of either Monel or stainless steel AISI 316L. KF flanges are connected using FFKM O-rings. A cold trap is located downstream of the reaction chamber, allowing for the deposition of the reaction products from the etching process. Liquid nitrogen is used as a frigorific agent. A constant and steady volume flow is ensured by the pumping unit at the very end of the line. A flow chart of the experimental setup is shown in Figure 2.

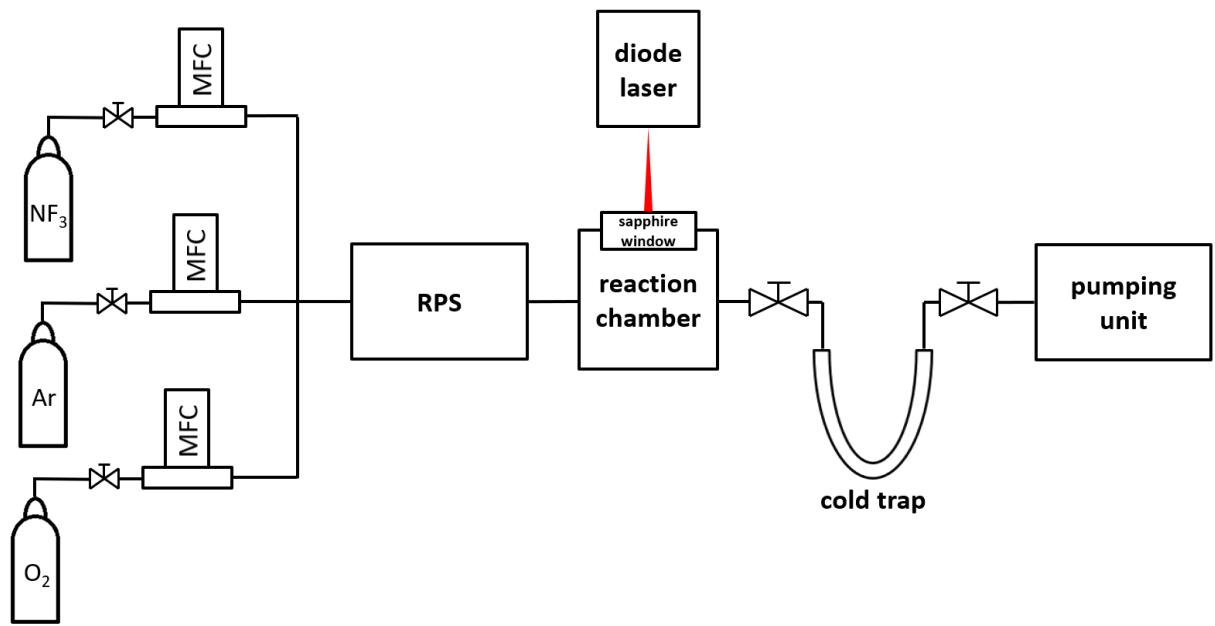

*Figure 2: Flow chart of the plasma line used for the experiments. Three different gases (Ar, NF$_3$ and O$_2$) can precisely be introduced into a Remote Plasma Source (RPS) using three Mass Flow Controllers (MFCs). O$_2$ and NF$_3$ thereby serve as etching agents and are therein radicalized. The substrate material is positioned in the reaction chamber and can be heated up to 1200°C using a laser heating system. The reaction products can then by deposited in a cold trap cooled with liquid nitrogen and further be analyzed.*

## 3. Experimental

During the experiments, both materials, PyC and SiC, were exposed to F/NF$_x$ and O radicals created in the RPS with the laser heating system set to different temperatures. Thereby, each compound was individually inserted into the reaction chamber of the plasma line on a Monel carrier.

### 3.1. Fluorination of pyrolytic carbon

The exact course of the fluorination of carbon depends on several parameters, including the physical form, in which it is exposed to the plasma as well as its origin and history in terms of production. For the fluorination experiments, one half of a disc of pyrolytic carbon with a diameter of 25.4 mm and a thickness of 6.35 mm served as a substitute material for the PyC layer of a TRISO particle. The PyC was manufactured by Kurt J. Lesker via chemical vapor deposition, had a purity of 99.999% and is usually being used as a sputtering target (see Figure 3a). The PyC substrate was exposed to fluorine radicals at temperature intervals of 400 °C, starting at room temperature up to 1200 °C. Exposure time was 30 min at room temperature, 30 min at 400 °C, 30 min at 800 °C, and 45 min at 1200 °C.

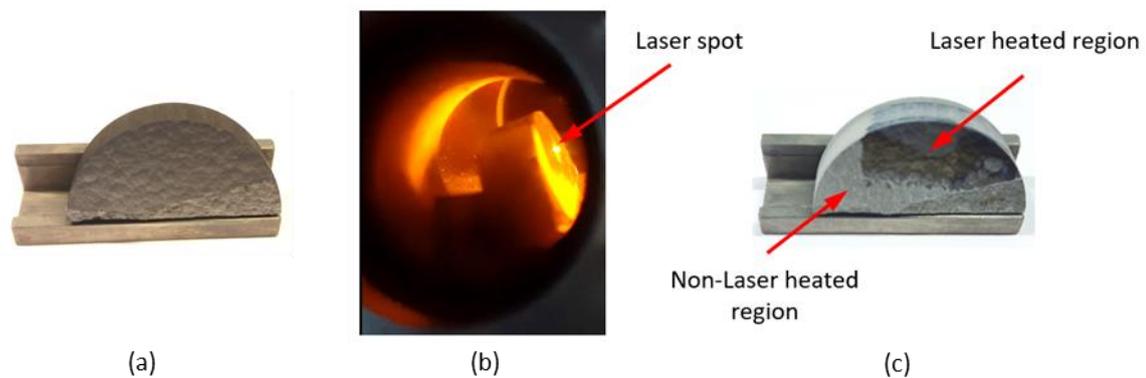

*Figure 3: PyC substrate on Monel carrier after being exposed to F radicals from NF$_3$ plasma for 30 minutes at room temperature without laser heating (a). The same sample in the reaction chamber at an NF$_3$ volume flow of 100 sccm and heated to 1200 °C; glimmering visible where the laser hits the sample (b). Same sample after fluorination for 45 minutes (c). Laser heating took place in the dark upper right region. The whiteish region is rich in fluorine, most likely graphite fluoride CF$_x$. The color thereby indicates x > 1.0.*

### 3.2. Oxygenation of pyrolytic carbon

The substrate material used for the oxygenation experiments of PyC was identical to the material used for the fluorination experiments (see Figure 4a). It was heated to a temperature of 1200 °C for 46 min using the laser heating system, showing bright glimmering and sparking, where the laser hit the sample (see Figure 4b). The volume flow of O$_2$ was set to 400 sccm.

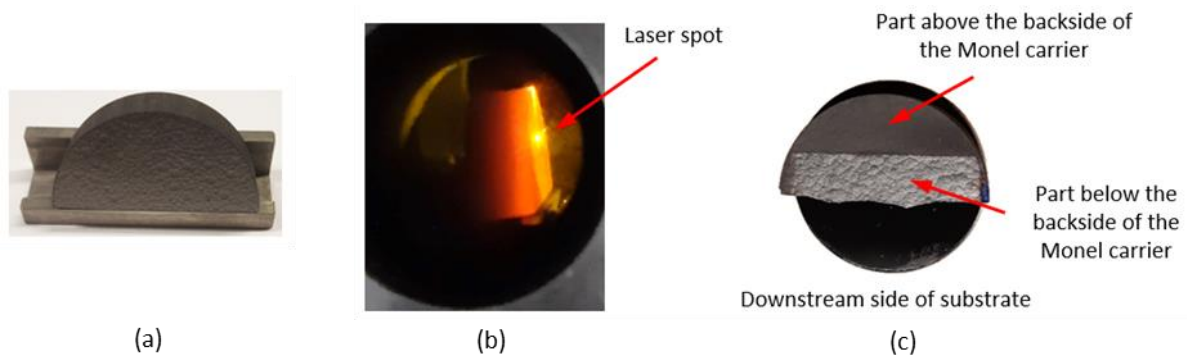

*Figure 4: Pyrolytic carbon substrate on Monel carrier before being exposed to oxygen radicals from O$_2$ plasma (a). The same sample in the reaction chamber at an O$_2$ volume flow of 400 sccm and heated to 1200 °C; glimmering and sparking visible where the laser hits the sample (b). Back side of the sample. Two clearly distinguishable zones are visible (c). The line separating these two indicates where the sample was in contact with the Monel carrier, thus separating the back side into a*

*part that was exposed to a high radical concentration (upper part) and a part that was in a shaded region and thus exposed to a lower radical concentration only (lower part).*

### 3.3. Fluorination of silicon carbide

The fluorination of SiC was performed on a sintered SiC sample with a square shape with a side length of 10 mm and a thickness of 2 mm (see Figure 5a). The SiC was supplied by Goodfellow and had a total mass of 663.4 mg. The sample was heated to 1200 °C during fluorination using the laser heating system, showing bright glimmering at the spot, where being hit by the laser (see Figure 5b). The volume flow of $NF_3$ was set to 100 sccm.

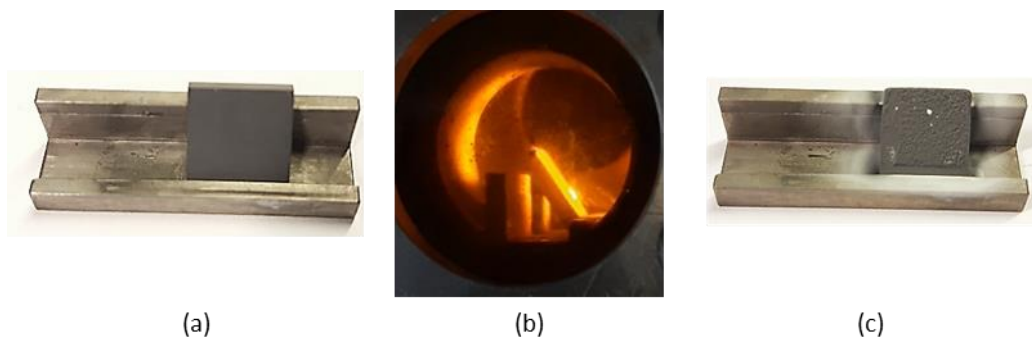

(a)　　　　　　　　　　　(b)　　　　　　　　　　　(c)

*Figure 5: Silicon carbide substrate on Monel carrier before being exposed to F/NF$_x$ radicals (a). The same sample in the reaction chamber at an NF$_3$ volume flow of 100 sccm and heated to 1200 °C; glimmering visible where the laser hits the sample (b). Same sample after fluorination for 30 minutes at 1200 °C (c). Laser heating covered almost the entire sample with the exception of region of about 1 mm width on the left as well as the lower side.*

### 3.4. Oxygenation of silicon carbide

Although SiC is known to be inert to oxygen under most conditions, it may undergo oxidation at higher temperatures and low oxygen partial pressures [10]. Thus, one oxygenation experiment was performed for a comparison to the SiC sample treated with F/NF$_x$ radicals. The heating temperature was 1200 °C, the $O_2$ volume flow was 100 sccm for 9 minutes and 400 sccm for an additional 30 minutes. During the exposure to oxygen radicals, the entire SiC sample was intensively glowing. Visual inspection of the sample after removal from the reaction chamber did not reveal any changes (see Figure 6a and 6c).

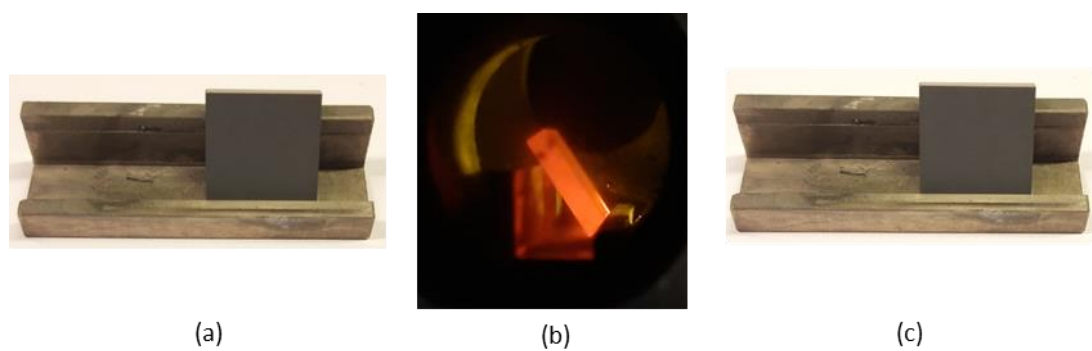

(a)　　　　　　　　　　　(b)　　　　　　　　　　　(c)

*Figure 6: Silicon carbide substrate on Monel carrier before being exposed to oxygen radicals (a). The same sample in the reaction chamber at an O$_2$ volume flow of 400 sccm and heated to 1200 °C; glimmering visible where the laser hits the sample (b). In contrast to the fluorination experiment, the entire sample is glowing, indicating homogeneous heating. Same sample after oxidation for 30 minutes at 1200 °C, visual inspection showing no difference in optical appearance (c).*

## 4. Results
### 4.1. Fluorination of pyrolytic carbon

As described in chapter 3.1, the fluorination of PyC took place at temperature intervals of 400 °C, starting at room temperature. Table 1 shows the respective mass difference of the substrate material after each trial compared to the incident mass. Note that the change of mass for room temperature and 400 °C are close to the measurement uncertainties of the used scale of ± 0.1 mg. To check, whether the change in mass is significant or not, additional tests will have to be performed. In any case, these differences do not influence the argumentation in this paper.

*Table 1: Mass differences of the pyrolytic carbon sample after being exposed to $F/NF_x$ radicals from $NF_3$ plasma at different temperatures. The same sample was used for different trials. Positive mass differences correspond to a gain in mass due to formation of graphite fluoride, negative mass differences to a loss in mass due to etching.*

| temperature [°C] | time [min] | mass difference [mg] |
|---|---|---|
| r.t. | 30 | + 0.18 |
| 400 | 30 | − 0.15 |
| 800 | 30 | + 2.45 |
| 1200 | 45 | − 4.78 |

Scanning electron microscopy (SEM) of the sample showed significant differences in the topography of the regions with and without laser heating (cf. Figure 3), respectively. Whereas the non-heated area appears very rough with a highly uneven structure (see Figure 7), the opposite holds for the area that was heated by the laser to 1200 °C (see Figure 8).

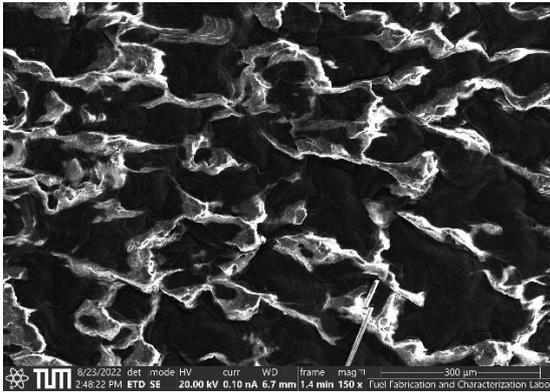

*Figure 7: SEM image of the whiteish region of the graphite substrate shown in Figure 3c, where the laser did not directly hit the sample (non-laser heated region).*

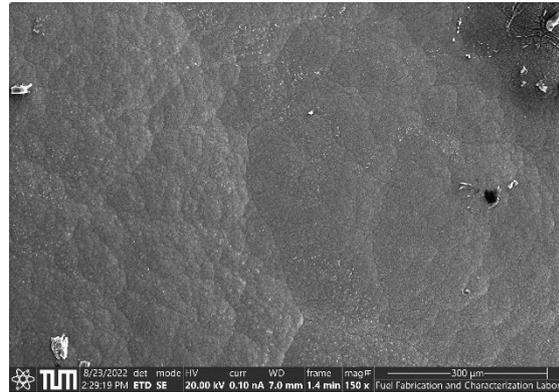

*Figure 8: SEM image of the dark region of the graphite substrate shown in Figure 3c, where the laser directly hit the sample (laser heated region).*

Figure 9a shows the transition area between the laser heated region and the non-heated region. The results of energy dispersive X-ray (EDX) mapping of the same area is shown in Figure 9b.

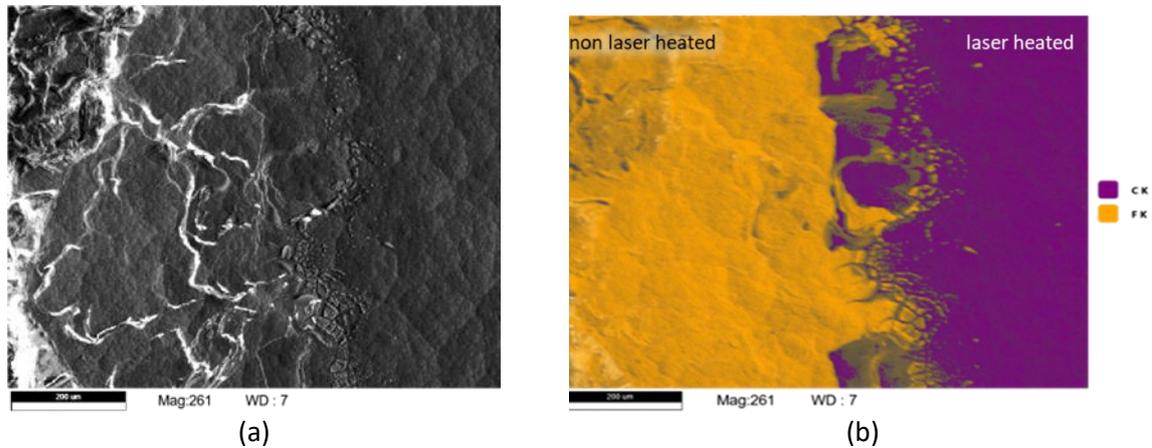

*Figure 9: (a) SEM image of the transition region of the part being heated by the laser and the region not directly irradiated with the laser. (b) Overlay of an EDX map on the SEM image from (a). The yellow region is rich in fluorine species whereas the purple region mainly shows carbon, i.e., graphite.*

### 4.2. Oxygenation of pyrolytic carbon

The difference in mass after oxygenation of the PyC for 46 min accounted for 30.9 mg. Figure 10 and Figure 11 show an SEM image of the graphite substrate before and after laser-heated oxygenation, respectively. The topography of the PyC surface shows only small changes, appearing somewhat smoother after oxygenation.

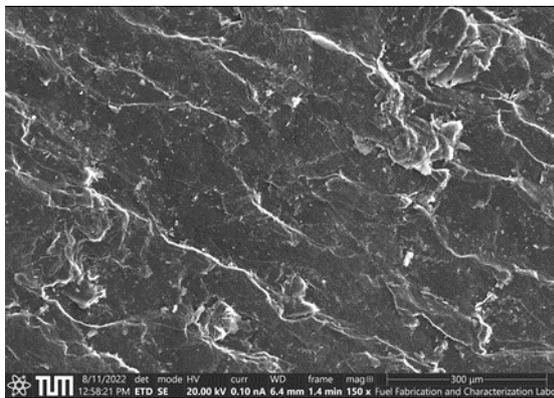

*Figure 10: SEM image of the graphite substrate shown in Figure 4a before plasma treatment.*

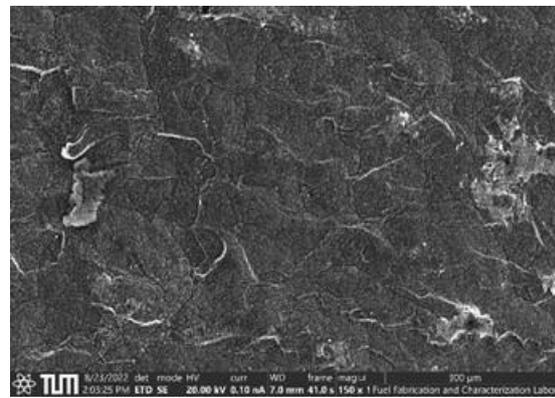

*Figure 11: SEM image of the upper region of the graphite substrate shown in Figure 4c after oxygen plasma exposure*

### 4.3. Fluorination of silicon carbide

The difference in mass after fluorination of the SiC sample for 30 min accounted for 322.8 mg. Visual inspection of the SiC sample shows clear signs of degradation. The sample has become thinner, its corners have been rounded and the surface appears significantly rougher compared to the initial state (cf. Figure 5a and 5c). In addition, the formation of white residues could be observed on the carrier around the sample. Again, the sample was also investigated using scanning electron microscopy prior and after plasma treatment. Figure 12 and Figure 13 show the surface of the SiC substrate before and after fluorination, respectively.

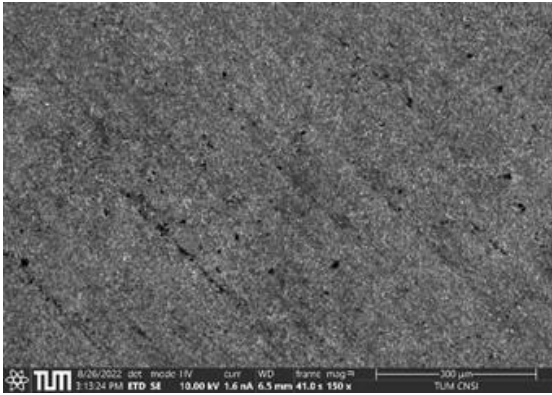 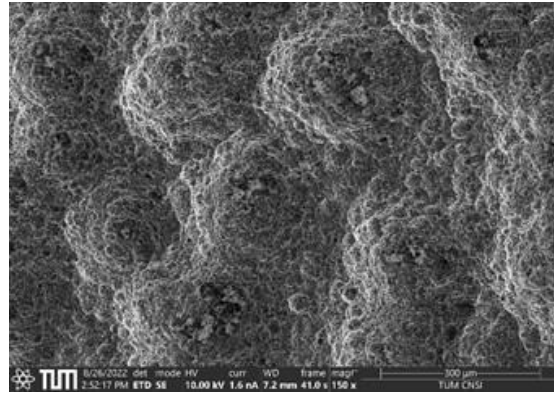

*Figure 12: SEM image of the SiC substrate shown in the left picture in Figure 5a before NF$_3$ plasma treatment.*

*Figure 13: SEM image of the SiC substrate shown in the right picture in Figure 5c after NF$_3$ plasma treatment.*

### 4.4. Oxygenation of silicon carbide

The difference in mass after oxygenation of the SiC sample for 40 min accounted for below 0.1 mg. On a microscopic scale, the surface of the substrate appears unchanged, as shown by SEM investigation. Figure 14 shows the surface of the SiC surface after oxygenation. For SiC before oxygenation, please refer to Figure 12.

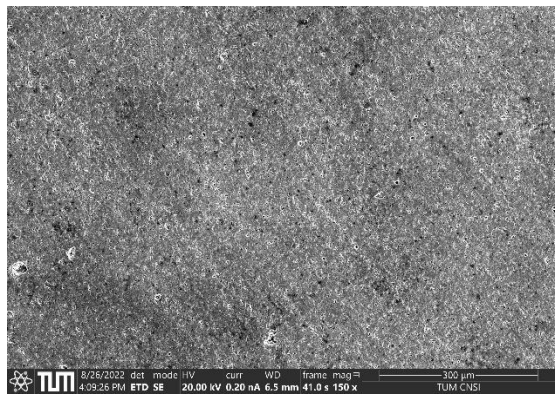

*Figure 14: SEM image of the SiC substrate shown in the right picture of Figure 6c after oxygen plasma treatment at 1200 °C for 30 minutes.*

In summary, SiC and PyC show quite a distinct etching behavior when exposed to F/NF$_x$ or oxygen radicals, respectively. Table 2 shows the summarized etching rates for both materials.

*Table 2: Summary of the etching rates of PyC and SiC using either NF$_3$ or molecular O$_2$ as a feed gas for radical generation. Experiments using NF$_3$ were performed at a volume flow of 100 sccm and a temperature of 1200 °C.*

|      | NF$_3$    | O$_2$     |
|------|-----------|-----------|
| **PyC**  | 6.4 mg/h   | 40.3 mg/h |
| **SiC**  | 645.5 mg/h | 0.1 mg/h  |

## 5. Discussion
### 5.1. Fluorination of pyrolytic carbon

A comparison of Figure 7 and Figure 8, both images acquired in secondary electron (SE) mode, shows a significant difference in the topography of the laser heated and the non-laser heated region, respectively. The rough and uneven surface of the non-heated region originates from the formation of CF$_x$, which serves as a passivation layer prohibiting further ablation and thus protecting the underlying graphite from being oxidized and leading to an increase in mass (see Table 1). The EDX map shown in Figure 9b underlines the interpretation. At the same time, according to Delabarre and coworkers [2],

$CF_x$ increases in thermal stability with increasing temperature up to 600 °C and decomposes at exceeding temperatures. Although this temperature cannot be confirmed as the sample showed an increase in mass even at 800 °C, the region laser-heated to 1200 °C does not show a passivation layer, but pure carbon instead. Thus, a rapid decomposition of $CF_x$ takes place at this temperature. This is reflected by the decrease in mass at this temperature (see Table 1). Due to the low etching rates, no analysis of the volatile reaction products was possible. However, for the laser heated region, the formation of carbon tetrafluoride $CF_4$ according to Equation 1

$$C + 4\,F \longrightarrow CF_4 \tag{1}$$

is very likely together with some $C_2F_6$. Further research is aimed at quantifying the reaction products through larger scale experiments.

### 5.2. Oxygenation of pyrolytic carbon

A comparison of Figure 10 and Figure 11, both images acquired in SE mode, shows a slightly smoother surface of the PyC sample after oxygenation. However, the change in appearance is by far not as significant as for the fluorination of the PyC samples. This seems plausible, as in the case of oxygenation, no protective layer is being formed. Instead, the formation of volatile CO and $CO_2$ according to Equations 2 and 3 exposes underlaying carbon layers, which are similar in optical appearance to the ones removed.

$$C + O \longrightarrow CO \tag{2}$$

$$C + 2\,O \longrightarrow CO_2 \tag{3}$$

### 5.3. Fluorination of silicon carbide

A comparison of Figure 12 and Figure 13, both images acquired in SE mode, shows a clear degradation of the sample after fluorination with an undulating, but nevertheless regular surface. This degradation originates from the SiC sample being heavily attacked by the $F/NF_x$ radicals, most likely reacting to $SiF_4$ and $CF_4$ according to Equations 4 and 5.

$$Si + 4\,F \longrightarrow SiF_4 \tag{4}$$

$$C + 4\,F \longrightarrow CF_4 \tag{5}$$

Both compounds, $SiF_4$ as well as $CF_4$, are highly volatile and readily enter the gas phase. They can thus be easily separated from the solid substrate material. The white residues accompanying the reaction are likely due to the formation of $CF_x$, and further research will quantify these byproducts.

### 5.4. Oxygenation of silicon carbide

A comparison of Figure 12 and Figure 14, both images acquired in SE mode, shows no significant difference of the sample before and after oxygenation. Thus, SiC does not show any degradation when exposed to oxygen radicals, which is additionally underlined by the missing difference in mass before and after the experiment.

# 6. Conclusion and Outlook
## 6.1. Conclusion

Four different sets of experiments were performed in order to investigate the etching behavior of F/NF$_x$ and oxygen radicals on this barrier system. Based on the data derived from these experiments, the removal of the PyC and SiC coating via a combined fluorination/oxygenation process seems to be a viable option for the treatment of TRISO fuel particles. For PyC, at 1200 °C the speed of the etching reaction is significantly higher for O$_2$ than for NF$_3$ by a factor of more than 6. For SiC, this difference is even more pronounced, although in favor of NF$_3$ over O$_2$. Here, the reaction speed differs by a factor of almost 6000, leaving SiC basically unaltered by O$_2$, but it being heavily attacked by NF$_3$. As SiC is not significantly attacked by O radicals even at a temperature of 1200 °C, the carbon layers can selectively be removed using an O plasma. SiC on the other hand can effectively be removed using F/NF$_x$ radicals from an NF$_3$ plasma. The graphite layer thereby forms a passivation CF$_x$ layer, protecting it from further fluorine attack if the temperature does not exceed 800 °C. At the same time, this layer also prevents the graphite from oxidation with O radicals. However, at higher temperatures this layer decomposes making the graphite susceptible once again to oxidation by O radicals.

## 6.2. Outlook

Independent of its deployment, the spent TRISO fuel from operating reactors and especially any future reprocessing of that fuel must be taken into account. Thereby, an additional waste stream segregation and volume reduction is highly desirable. While these first tests were successful in demonstrating the potential for chemical processing, two key issues for further research were identified:

- There is a need to demonstrate the layer-by-layer processing in an actual TRISO arrangement. We are planning on fabricating TRISO with depleted or natural uranium cores to demonstrate the processing on a physically similar system.
- Activation products of the protective layer system such as $^{14}$C form volatile compounds that must be captured and reprocessed, including $^{14}$CO$_2$.

Although a significant amount of additional research is necessary, the experiments have demonstrated that plasma-aided chemical processing based on the use of oxygen and F/NF$_x$ radicals may provide a safe and convenient avenue for TRISO fuel processing not previously explored.

**Acknowledgements**

Work at the Technische Universität München was supported through a combined grant (FRM2023) from the Bundesministerium für Bildung und Forschung (BMBF), Germany and the Bayerisches Staatsministerium für Wissenschaft und Kunst (StMWK). The work at Philipps-Universität Marburg was supported by the Deutsche Forschungsgemeinschaft through the grant DFG KR3595/10-1.